\documentclass[journal]{IEEEtran}
\ifCLASSINFOpdf
\else
\fi

\usepackage{cite}
\pdfoutput=1

\usepackage{graphicx}
\graphicspath{{./Figures/}}
\DeclareGraphicsExtensions{.pdf,.jpg,.png}

\usepackage{mathrsfs,amsmath}
\usepackage{mathtools}
\usepackage{algorithm}
\usepackage{algpseudocode}

\usepackage{amsfonts}
\usepackage{tabularx}
\usepackage{multicol}
\usepackage{multirow}
\usepackage{array}
\usepackage{enumitem}
\usepackage{placeins}
\usepackage{hhline}
\usepackage[bottom]{footmisc}

\ifCLASSOPTIONcompsoc
 \usepackage[caption=false,font=normalsize,labelfont=sf,textfont=sf]{subfig}
\else
 \usepackage[caption=false,font=footnotesize]{subfig}
\fi

\makeatletter
\newcommand{\multiline}[1]{%
  \begin{tabularx}{\dimexpr\linewidth-\ALG@thistlm}[t]{@{}X@{}}
    #1
  \end{tabularx}
}
\makeatother

\begin{document}

\title{GANPOP: Generative Adversarial Network Prediction of Optical Properties from Single Snapshot Wide-field Images}

\author{Mason~T.~Chen,
        Faisal~Mahmood,
        Jordan~A.~Sweer,
        and~Nicholas~J.~Durr~% 
\thanks{All authors are with the Department
of Biomedical Engineering, Johns Hopkins University, Baltimore,
MD, 21218. Contact e-mail: ndurr@jhu.edu.}% 
% \thanks{Manuscript received Month xx, 2019; revised Month xx, 2019.}
}

\maketitle

\begin{abstract}
We present a deep learning framework for wide-field, content-aware estimation of absorption and scattering coefficients of tissues, called Generative Adversarial Network Prediction of Optical Properties (GANPOP). Spatial frequency domain imaging is used to obtain ground-truth optical properties from \textit{in vivo} human hands, freshly resected human esophagectomy samples and homogeneous tissue phantoms. Images of objects with either flat-field or structured illumination are paired with registered optical property maps and are used to train conditional generative adversarial networks that estimate optical properties from a single input image. We benchmark this approach by comparing GANPOP to a single-snapshot optical property (SSOP) technique, using a normalized mean absolute error (NMAE) metric. In human gastrointestinal specimens, GANPOP estimates both reduced scattering and absorption coefficients at 660 nm from a single 0.2 mm\textsuperscript{-1} spatial frequency illumination image with 58\% higher accuracy than SSOP. When applied to both \textit{in vivo} and \textit{ex vivo} swine tissues, a GANPOP model trained solely on human specimens and phantoms estimates optical properties with approximately 43\% improvement over SSOP, indicating adaptability to sample variety. Moreover, we demonstrate that GANPOP estimates optical properties from flat-field illumination images with similar error to SSOP, which requires structured-illumination. Given a training set that appropriately spans the target domain, GANPOP has the potential to enable rapid and accurate wide-field measurements of optical properties, even from conventional imaging systems with flat-field illumination.   
\end{abstract}

\begin{IEEEkeywords}
optical imaging, tissue optical properties, neural networks, machine learning, spatial frequency domain imaging
\end{IEEEkeywords}

\IEEEpeerreviewmaketitle

\section{Introduction}
\IEEEPARstart{T}{he} optical properties of tissues, including the absorption ($\mu_{a}$) and reduced scattering ($\mu_{s}'$) coefficients, can be useful clinical biomarkers for measuring trends and detecting abnormalities in tissue metabolism, tissue oxygenation, and cellular proliferation \cite{doi:10.1146/annurev.physchem.47.1.555,drezek2003light,maloney2018review,mourant2000light,Steelman:19}. Optical properties can also be used for contrast in functional or structural imaging \cite{Lin2011,Shah4420}. Thus, quantitative imaging of tissue optical properties can facilitate more objective, precise, and optimized management of patients. 

To measure optical properties, it is generally necessary to decouple the effects of scattering and absorption, which both influence the measured intensity of remitted light. Separation of these parameters can be achieved with temporally or spatially resolved techniques, which can each be performed with measurements in the real or frequency domains. Spatial Frequency Domain Imaging (SFDI) decouples absorption from scattering by characterizing the tissue modulation transfer function to spatially modulated light \cite{dognitz1998determination, cuccia_quantitation_2009}. This approach has significant advantages in that it can easily be implemented with a consumer grade camera and projector, and achieve rapid, non-contact mapping of optical properties. These advantages make SFDI well-suited for applications that benefit from wide-field characterization of tissues, such as image-guided surgery \cite{pharaon_early_2010, gioux_first--human_2011} and wound characterization \cite{kaiser_noninvasive_2011,WEINKAUF2019555, doi:10.1002/lsm.22692}. Additionally, recent work has explored the use of SFDI for improving endoscopic procedures \cite{angelo_real-time_2017,nandy_label-free_2018}.

\begin{figure}[t]
\centering
\includegraphics[width=3.5in]{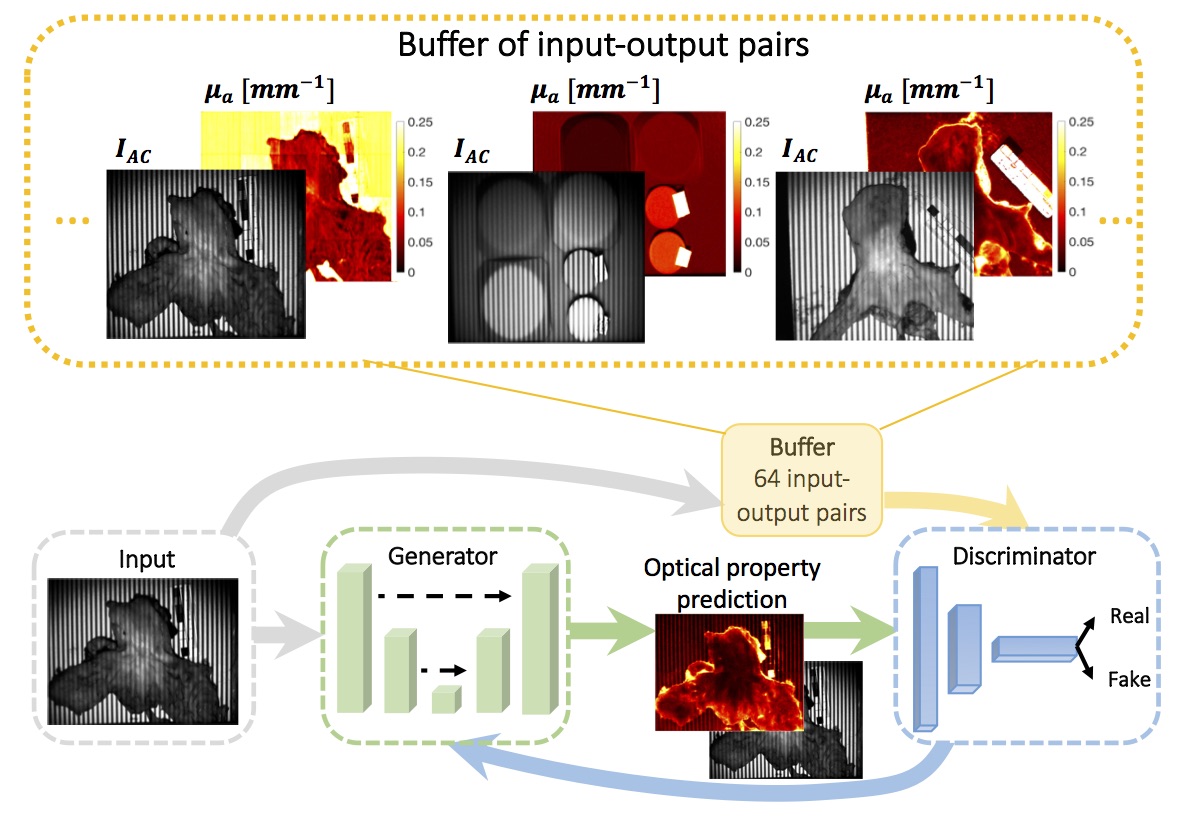}
\caption{Proposed conditional Generative Adversarial Network (cGAN) architecture. The generator is a combination of ResNet and U-Net and is trained to produce optical property maps that closely resemble SFDI output. The discriminator is a three-layer classifier that operates on a patch level and is trained to classify the output of the generator as ground truth (real) or generated (fake). The discriminator is updated using a history of 64 previously-generated image pairs.}
\label{fig:method_network}
\end{figure}

Although SFDI is finding a growing number of clinical applications, there are remaining technical challenges that limit its adoption. First, SFDI requires structured light projection with carefully-controlled working distance and calibration, which is especially challenging in an endoscopic setting. Second, it is difficult to achieve real-time measurements. Conventional SFDI requires a minimum of six images per wavelength (three distinct spatial phases at two spatial frequencies) to generate a single optical property map. A lookup table (LUT) search is then performed for optical property fitting. The recent development of real-time single snapshot imaging of optical properties (SSOP) has reduced the number of images required per wavelength from 6 to 1, considerably shortening acquisition time \cite{vervandier_single_2013}. However, SSOP introduces image artifacts arising from single-phase projection and frequency filtering, which corrupt the optical property estimations. To reduce barriers to clinical translation, there is a need for optical property mapping approaches that are simultaneously fast and accurate while requiring minimal modifications to conventional camera systems.

Here, we introduce a deep learning framework to predict optical properties directly from single images. Deep networks, especially convolutional neural networks (CNNs), are growing in popularity for medical imaging tasks, including computer-aided detection, segmentation, and image analysis\cite{suzuki_overview_2017,7404017,7426826}. We pose the optical property estimation challenge as an image-to-image translation task and employ generative adversarial networks (GANs) to efficiently learn a transformation that is robust to input variety. First proposed in \cite{goodfellow_generative_2014}, GANs have improved upon the performance of CNNs in image generation by including both a generator and a discriminator. The former is trained to produce realistic output, while the latter is tasked to classify generator output as real or fake. The two components are trained simultaneously to outperform each other, and the discriminator is discarded once the generator has been trained. When both components observe the same type of data, such as text labels or input images, the GAN model becomes conditional. Conditional GANs (cGANs) are capable of making structured predictions by incorporating non-local, high-level information. Moreover, because they can automatically learn a loss function instead of using a handcrafted one, cGANs have the potential to be an effective and generalizable solution to various image-to-image translation tasks \cite{mirza2014conditional,Isola_2017}. In medical imaging, cGANs have been proven successful in many applications, such as image synthesis \cite{8662628}, noise reduction \cite{8340157}, and sparse reconstruction \cite{8233175}. In this study, we train cGAN networks on a series of structured or flat-field illumination images paired with corresponding optical property maps (Fig. \ref{fig:method_network}). We demonstrate that the GANPOP approach produces rapid and accurate estimation from input images from a wide variety of tissues using a relatively small set of training data. 

\section{Related Work}
\subsection{Diffuse reflectance imaging}
Optical absorption and reduced scattering coefficients can be measured using temporally or spatially resolved diffuse reflectance imaging. Approaches that rely on point illumination inherently have a limited field of view \cite{Swartling:03,doi:10.1063/1.3116135}. Non-contact, hyperspectral imaging techniques measure the attenuation of light at different wavelengths, from which the concentrations of tissue chromophores, such as oxy- and deoxy-hemoglobin, water, and lipids, can be quantified \cite{palmer_quantitative_2009}. A recent study has also proposed using a Bayesian framework to infer tissue oxygen concentration by recovering intrinsic multispectral measurements from RGB images \cite{7859372}. However, these methods fail to unambiguously separate absorption and scattering coefficients, which poses a challenge for precise chromophore measurements. Moreover, accurate determination of both parameters is critical for the detection and diagnosis of diseases \cite{ doi:10.1146/annurev.physchem.47.1.555, Steelman:19 }.

\subsection{Single snapshot imaging of optical properties}
SSOP achieves optical property mapping from a single structured light image. Using Fourier domain filtering, this method separates DC (planar) and AC (spatially modulated) components from a single-phase structured illumination image \cite{vervandier_single_2013}. A grid pattern can also be applied to simultaneously extract optical properties and three-dimensional profile measurements \cite{van_de_giessen_real-time_2015}. When tested on homogeneous tissue-mimicking phantoms, this method is able to recover optical properties within 12\% for absorption and 6\% for reduced scattering using conventional profilometry-corrected SFDI as ground truth.

\subsection{Machine learning in optical property estimation}
Despite its prevalence and increasing importance in the field of medical imaging, machine learning has only recently been explored for optical property mapping. This includes a random forest regressor to replace the nonlinear model inversion \cite{ml_op}, and using deep neural networks to reconstruct optical properties from multifrequency measurements \cite{zhao2018deep}. Both of these approaches aim to bypass the time-consuming LUT step in SFDI. However, they require diffuse reflectance measurements from multiple images to achieve accurate results and consider each pixel independently.

\section{Contributions}
We propose an adversarial framework for learning a content-aware transformation from single illumination images to optical property maps. In this work, we:

\begin{enumerate}[label=\arabic*)]
\item develop a data-driven model to estimate optical properties directly from input reflectance images;
\item demonstrate advantages of structured versus flat-field light as an input to determine optical properties;
\item perform cross-validated experiments, comparing our technique with model-based SSOP and other deep learning-based methods; and
\item acquire and make publicly-available a dataset of registered flat-field-illumination images, structured-illumination images, and ground-truth optical properties of a variety of \textit{ex vivo} and \textit{in vivo} tissues.
\end{enumerate}

\section{Methods}
For training and testing of the GANPOP model, single structured or flat-field illumination images were used, paired with registered optical property maps. To generate ground truth optical properties, conventional six-image SFDI was implemented. GANPOP performance was analyzed and compared to other techniques both in unseen tissues of the same type as the training data (new \textit{ex vivo} esophagus) and in different tissue types (\textit{in vivo} and \textit{ex vivo} swine gastrointestinal tissues). 

\subsection{Hardware}
\begin{figure}[!htb]
\centering
\includegraphics[width=3.5in]{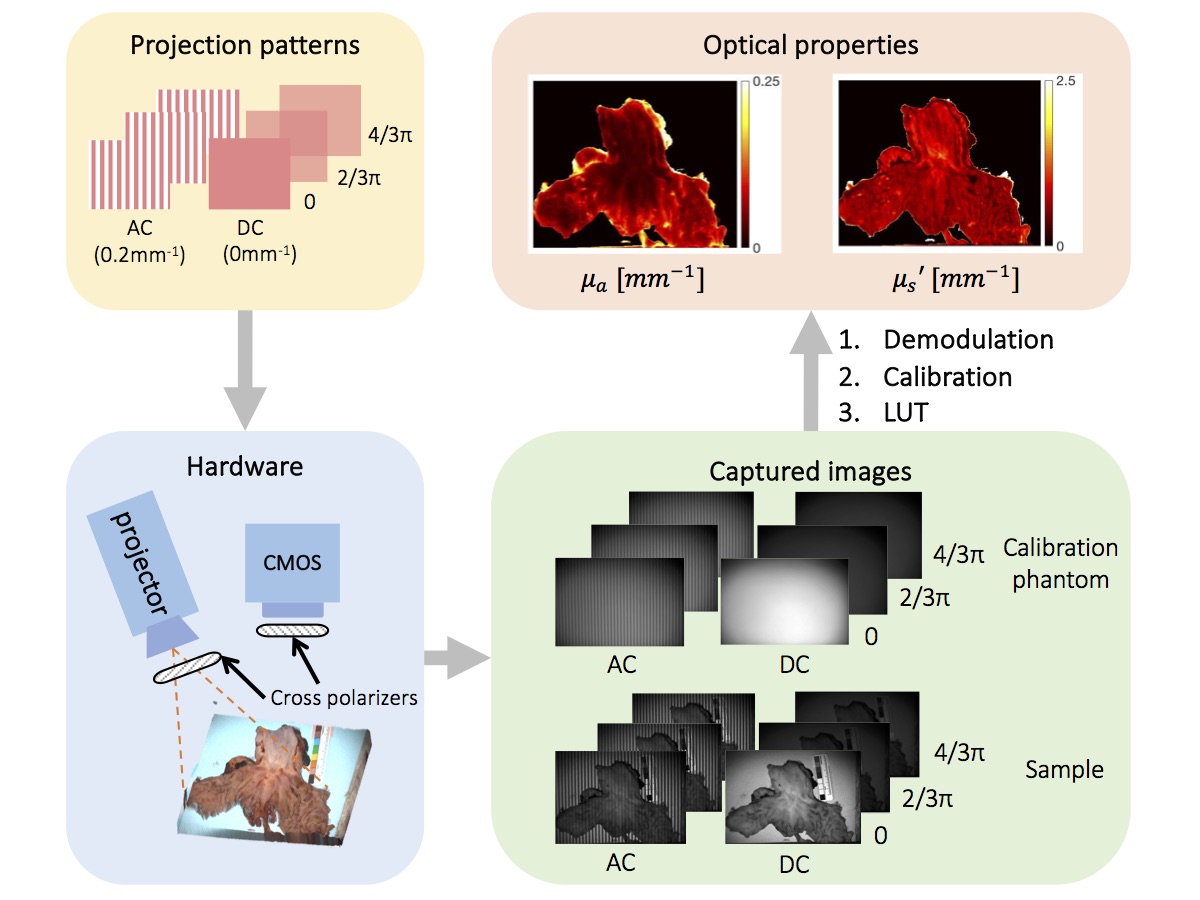}
\caption{Overview of conventional SFDI illumination patterns, hardware, and processing flow. SFDI captures six frames (three phase offsets at two different spatial frequencies) to generate an absorption and reduced scattering map. DC stands for planar illumination images and AC stands for spatially modulated images. To calculate optical properties, acquired data is demodulated, calibrated against a reference phantom, and inverted using a lookup table.}
\label{fig:method_SFDI}
\end{figure}

In this study, all images were captured using a commercially available SFDI system (Reflect RS\textsuperscript{TM}, Modulated Imaging Inc.). A schematic of the system is shown in Fig. \ref{fig:method_SFDI}. Cross polarizers were utilized to reduce the effect of specular reflections, and images were acquired in a custom-built light enclosure to minimize ambient light. Raw images, after 2x2 pixel hardware binning, were 520 $\times$ 696 pixels, with a pixels size of 0.278 mm in the object space. 

\subsection{SFDI ground truth optical properties}
Ground truth optical property maps were generated using conventional SFDI with 660 nm light following the method from Cuccia \textit{et al.} \cite{cuccia_quantitation_2009}. First, images of a calibration phantom with homogeneous optical properties and the tissue of interest are captured under spatially modulated light. We used a flat polydimethylsiloxane-titanium dioxide (PDMS-TiO\textsubscript{2}) phantom with reduced scattering coefficient of 0.957 mm\textsuperscript{-1} and absorption coefficient of 0.0239 mm\textsuperscript{-1} at 660 nm. We project spatial frequencies of 0 mm\textsuperscript{-1} (DC) and 0.2 mm\textsuperscript{-1} (AC), each at three different phase offsets ($0$, $\frac{2}{3}\pi$, and $\frac{4}{3}\pi$) for this study. AC images are demodulated at each pixel $x$ using: 

\begin{equation}
M_{AC}(x)=\frac{\sqrt{2}}{3}\cdot
\sqrt{
\begin{aligned}
(I_{1}(x)-I_{2}(x))^{2} & +(I_{2}(x)-I_{3}(x))^{2} \\
& +(I_{3}(x)-I_{1}(x))^{2},
\end{aligned}
}
\label{eq:demod}
\end{equation}
where $I_{1}$, $I_{2}$, and $I_{3}$ represent images at the three phase offsets. The spatially varying DC amplitude is calculated as the average of the three DC images. Diffuse reflectance at each pixel $x$ is then computed as:

\begin{equation}
R_{d}(x)=\frac{M_{AC}(x)}{M_{AC,ref}(x)}\cdot R_{d,predicted}
\label{eq:reflectance}.
\end{equation}

Here, $M_{AC,ref}$ denotes the demodulated AC amplitude of the reference phantom, and $R_{d,predicted}$ is the diffuse reflectance predicted by Monte Carlo models. Where indicated, we corrected for height and surface angle variation of each pixel from depth maps measured via profilometry. Profilometry measurements were obtained by projecting a spatial frequency of 0.15 mm\textsuperscript{-1} and calculating depth at each pixel \cite{gioux_three-dimensional_2009}. Finally, $\mu_{a}$ and $\mu_{s}'$ are estimated by fitting $R_{d,0mm^{-1}}$ and $R_{d,0.2mm^{-1}}$ into an LUT previously created using Monte Carlo simulations \cite{martinelli_analysis_2011}.

\subsection{Single Snapshot Optical Properties (SSOP)}
SSOP was implemented as the model-based alternative of GANPOP. This method separates DC and AC components from a single-phase structured-light image by frequency filtering with a 2D band-stop filter and a high-pass filter \cite{vervandier_single_2013}. Both filters are rectangular windows that isolate the frequency range of interest while preserving high-frequency content of the image. In this study, cutoff frequencies $f_{DC}=$ [0.16 mm\textsuperscript{-1}, 0.24 mm\textsuperscript{-1}] and $f_{AC}=$ [0, 0.16 mm\textsuperscript{-1}] were selected \cite{van_de_giessen_real-time_2015}. $M_{DC}$ can subsequently be recovered through a 2D inverse Fourier transform, and the AC component is obtained using an additional Hilbert transform. 

\begin{figure}[!htb]
\centering{}
\includegraphics[width=3.5in]{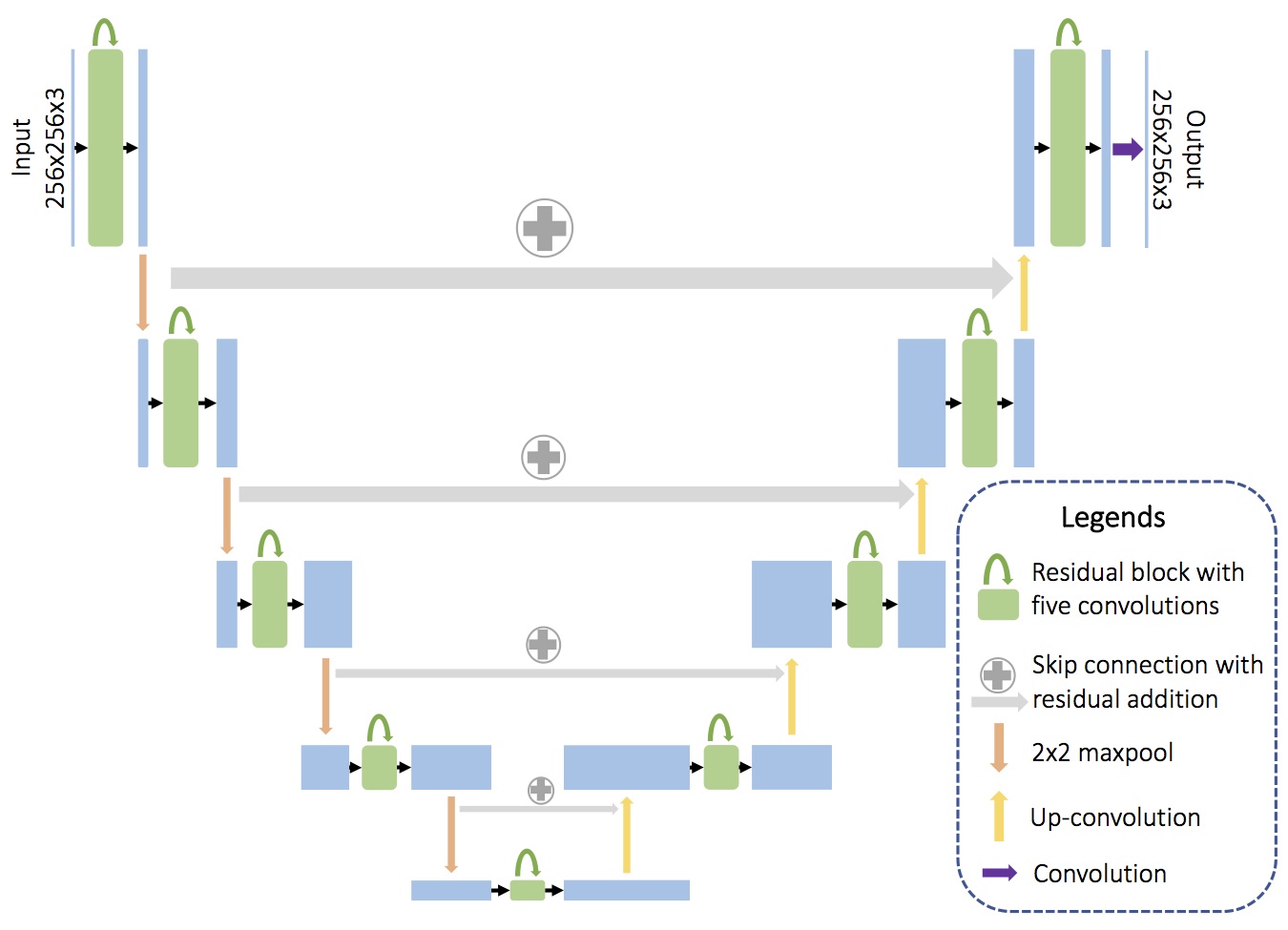}
\caption{Detailed architecture of the proposed generator. We use a fusion network that combines properties from ResNet and U-Net, including both short and long skip connections in the form of feature addition. Each residual block contains five convolution layers, with short skips between the first and the fifth layer.}
\label{fig:network}
\end{figure}

\subsection{GANPOP Architecture}
The GANPOP architecture is based on an adversarial training framework. When used in a conditional GAN-based image-to-image translation setup, this framework has the ability to learn a loss function while avoiding the uncertainty inherent in using hand-crafted loss functions \cite{Isola_2017,chen2018rethinking}. The generator is tasked with predicting pixel-wise optical properties from SFDI images while the discriminator classifies pairs of SFDI images and optical property maps as being real or fake (Fig. \ref{fig:method_network}). The discriminator additionally gives feedback to the generator over the course of training. The generator employs a modified U-Net consisting of an encoder and a decoder with skip connections \cite{Ronneberger_2015}. However, unlike the original U-Net, the GANPOP network includes properties of a ResNet, including short skip connections within each level \cite{He_2016} (Fig. \ref{fig:network}). Each residual block is a 3-layer building block with an additional convolutional layer on both sides. This ensures that the number of input features matches that of the residual block and that the network is symmetric \cite{Quan2016FusionNetAD}. Moreover, GANPOP generator replaces the U-Net concatenation step with feature addition, making it a fully residual network. Using $n$ as the total number of layers in the encoder-decoder network and $i$ as the current layer, long skip connections are added between the $i^{th}$ and the $(n-i)^{th}$ layer in order to sum features from the two levels. After the last layer in the decoder, a final convolution is applied to shrink the number of output channels and is followed by a $Tanh$ function. Regular ReLUs are used for the decoder and leaky ReLUs (slope = 0.2) for the encoder. We chose a receptive field of 70 $\times$ 70 pixels for our discriminator because this window captures two periods of AC illumination in each direction. This discriminator is a three-layer classifier with leaky ReLUs (slope = 0.2), as discussed in \cite{Isola_2017}. The discriminator makes classification decisions based on the current batch as well as a batch randomly sampled from 64 previously generated image pairs. Both networks are trained iteratively and the training process is stabilized by incorporating spectral normalization in both the generator and the discriminator \cite{miyato2018cgans}. The conditional GAN objective for generating optical property maps from input images ($G: X \rightarrow Y$) is:

\begin{equation}
  \begin{aligned}
\mathcal{L}_{\text{GAN}}(G,D) = & \mathbb{E}_{x,y\sim p_{\text{data}}(x,y)}[(D(x,y)-1)^2]\\
& +\mathbb{E}_{x\sim p_{\text{data}}(x)}[D(x,G(x))^2],
   \end{aligned}  
   \label{eq:loss_cgan}
 \end{equation}
where $G$ is the generator, $D$ the discriminator, and $p_{\text{data}}$ is the optimal distribution of the data. We empirically found that a least squares GAN (LSGAN) objective \cite{Mao_2017} produced slightly better performance in predicting optical properties than a traditional GAN objective \cite{goodfellow_generative_2014}, and so we utilize LSGAN in the networks presented here. An additional $\mathcal{L\textsubscript{1}}$ loss term was added to the GAN loss to further minimize the distance from the ground truth distribution and stabilize adversarial training:

 \begin{equation}
\mathcal{L}_{1}(S) = \mathbb{E}_{x,y\sim p_{\text{data}}}(x,y)[||y-G(x)||_1].
\label{eq:loss_l1}
 \end{equation}

The full objective can be expressed as:

 \begin{equation}
  \begin{aligned}
 \mathcal{L}(G,D) = \mathcal{L}_{\text{GAN}}(G, D) +  \lambda\mathcal{L}_{1}(G), 
   \end{aligned}
   \label{eq:loss_full}  
 \end{equation}
where $\lambda$ is the regularization parameter of the $\mathcal{L\textsubscript{1}}$ loss term. This optimization problem was solved using an \textit{Adam} solver with a batch size of 1 \cite{kingma2014adam}. The training code was implemented using Pytorch 1.0 on Ubuntu 16.04 with Google Cloud. For all experiments, $\lambda$ was set to 60. A total of 200 epochs was used with a learning rate of 0.0002 for half of the epochs and the learning rate was linearly decayed for the remaining half. Both networks were initialized from a Gaussian distribution with a mean and standard deviation of 0 and 0.02, respectively.

Conventional neural networks typically operate on three-channel (or RGB) images as input and output, with each channel representing red (R), green (G), or blue (B). In this study, four separate networks ($\text{N}_1$ to $\text{N}_4$) were trained for image-to-image translation with a variety of input and output parameters, summarized in Table \ref{table:networks}.

\begin{table}
\renewcommand{\arraystretch}{2.5}
\newcolumntype{L}{>{\centering\arraybackslash}m{0.15\linewidth}} 
\caption{Summary of networks trained in this study.}
\label{table:networks}
\centering

\begin{tabular}{p{0.02\linewidth}|LL|LL}
\hline
{} & \multicolumn{2}{c}{Input} & \multicolumn{2}{c}{Output}\\
\hline
$\text{N}_{i}$ & R channel & G channel & R channel & G channel\\
\hline
$\text{N}_{1}$ & $\frac{I_{AC}}{M_{DC,ref}}$ & $\frac{I_{AC}}{M_{AC,ref}}$ & $\mu_{a}$ & $\mu_{s}'$\\
$\text{N}_{2}$ & $\frac{I_{AC}}{M_{DC,ref}}$ & $\frac{I_{AC}}{M_{AC,ref}}$ & $\mu_{a,prof}$ & $\mu_{s,prof}'$\\
$\text{N}_{3}$ & $\frac{I_{DC}}{M_{DC,ref}}$ & $\frac{I_{DC}}{M_{AC,ref}}$ & $\mu_{a}$ & $\mu_{s}'$\\
$\text{N}_{4}$ & $\frac{I_{DC}}{M_{DC,ref}}$ & $\frac{I_{DC}}{M_{AC,ref}}$ & $\mu_{a,prof}$ & $\mu_{s,prof}'$\\
\hline
\end{tabular}

\end{table}

\begin{figure}
\centering
\includegraphics[width=3.5in]{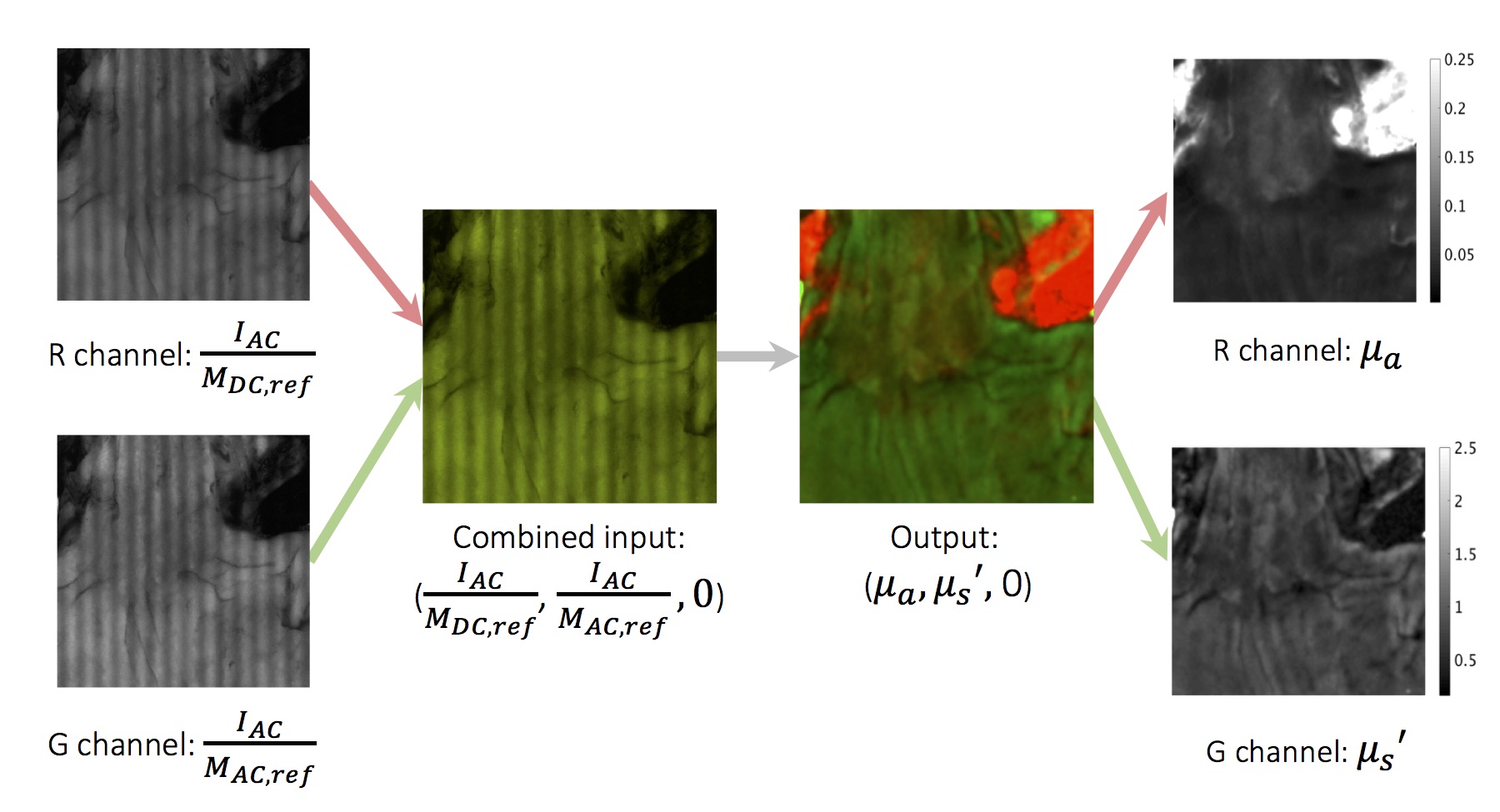}
\caption{Example input-output pair used in $N_{1}$ showing each individual channel as well as the combined RGB images. R and G channel of the output contain $\mu_{a}$ and $\mu_{s}'$, respectively. Thus, a high absorption appears red while a high scattering appears green.}
\label{fig:example_patch}
\end{figure}

For input, $I_{AC}$ and $I_{DC}$ represent single-phase raw images at 0.2 mm\textsuperscript{-1} and 0 spatial frequency, respectively. $M_{DC,ref}$ and $M_{AC,ref}$ are the demodulated DC and AC amplitude of the calibration phantom. Blue channels are left as zeros in all cases. It is important to note that $M_{AC,ref}$ and $M_{DC,ref}$ are measured only once during calibration before the imaging experiment and thus do not add to the total acquisition time. The purpose of these two terms is to account for drift of the system over time and correct for non-uniform illumination, making the patch used in the network origin-independent. These two calibration images are also required by traditional SFDI and the SSOP approaches. A network without calibration was empirically trained, and it produced 230\% and 58\% larger error than with calibration in absorption and scattering coefficients, respectively. A single output image contains both $\mu_{a}$ and $\mu_{s}'$ in different channels. Two dedicated networks were empirically trained for estimating $\mu_{a}$ and $\mu_{s}'$ independently, but no accuracy benefits were observed. Optical property maps calculated by non-profile-corrected SFDI were used as ground truth for $N_{1}$ and $N_{3}$. We also assessed the ability of GANPOP to learn both optical property estimation and sample height and surface normal correction by training and testing with profilometry-corrected data ($N_{2}$ and $N_{4}$).

All optical property maps for training and testing were normalized to have a consistent representation in the 8-bit images commonly used in CNNs \cite{vanhoucke_improving_2011}. We defined the maximum value of 255 to be 0.25 mm\textsuperscript{-1} for $\mu_{a}$ and 2.5 mm\textsuperscript{-1} for $\mu_{s}'$. Additionally, each image of size 520 $\times$ 696 was segmented at a random stride size into multiple patches of 256 $\times$ 256 pixels and paired with a registered optical property patch for training, as shown in Fig. \ref{fig:example_patch}.

% \begin{algorithm}
% \label{algo}
% \caption{}
% \label{euclid}
% \begin{flushleft}
%         \textbf{INPUT:} $N_i$ 256 $\times$ 256 patches of $I_{AC}$, $M_{DC,ref}$, $M_{AC,ref}$;\\
% 		\hspace*{3.35em} Trained GANPOP model
% \end{flushleft}
% \begin{algorithmic}[1]
% \For{$n_i=1,2,3...N_i$}
% \State \multiline{Assemble input $(C1, C2, 0)$, where:\\
% \hskip\algorithmicindent$C1=I_{AC}/M_{DC,ref}*255$, and \\
% \hskip\algorithmicindent$C2=I_{AC}/M_{AC,ref}*255$}
% \State{Forward propagation through GANPOP model} 
% \State \multiline{Recover optical properties from output $(O1, O2, 0)$:\\
% \hskip\algorithmicindent$\mu_{a}=O1/255*0.25$, and \\
% \hskip\algorithmicindent$\mu_{s}'=O2/255*2.5$}
% \State{Stitch output patches back to original size}
% \EndFor
% \end{algorithmic}
% \begin{flushleft}
%         \textbf{OUTPUT:} $\mu_{a}$ and $\mu_{s}'$ maps in mm\textsuperscript{-1}
% \end{flushleft}
% \end{algorithm}

\subsection{Tissue Samples}

\subsubsection{\textit{Ex vivo} human esophagus}
Eight \textit{ex vivo} human esophagectomy samples were imaged at Johns Hopkins Hospital for training and testing of our networks. All patients were diagnosed with esophageal adenocarcinoma and were scheduled for an esophagectomy. The research protocol was approved by the Johns Hopkins Institutional Review Board and consents were acquired from all patients prior to each study. All samples were handled by a trained pathologist and imaged within one hour after resection \cite{sweer2019wide}. 

Example raw images of a specimen captured by the SFDI system are shown in Fig. \ref{fig:method_SFDI}, \ref{fig:example_patch}, and \ref{fig:results}(a). All samples consisted of the distal esophagus, the gastroesophageal junction, and the proximal stomach. The samples contain complex topography and a relatively wide range of optical properties (0.02-0.15 mm\textsuperscript{-1} for $\mu_{a}$ and 0.1-1.5 mm\textsuperscript{-1} for $\mu_{s}'$ at $\lambda$ = 660 nm), making it suitable for training a generalizable model that can be applied to other tissues with non-uniform surface profiles. An illumination wavelength of 660 nm was chosen because it is close to the optimal wavelength for accurate tissue oxygenation measurements \cite{mazhar_wavelength_2010}.

In this study, six \textit{ex vivo} human esophagus samples were used for training of the GANPOP model and two used for testing. A leave-two-out cross validation method was implemented, resulting in four iterations of training for each network. Performance results reported here are from an average of these four iterations.

\subsubsection{Homogeneous phantoms}
The four GANPOP networks were also trained on a set of tissue-mimicking silicone phantoms made from PDMS-TiO\textsubscript{2} (P4, Eager Plastics Inc.) mixed with India ink as absorbing agent \cite{ayers_fabrication_2008}. To ensure homogeneous optical properties, the mixture was thoroughly combined and poured into a flat mold for curing. In total, 18 phantoms with unique combinations of $\mu_{a}$ and $\mu_{s}'$ were fabricated, and their optical properties are summarized in Fig. \ref{fig:ssop_phantoms}. 

% \begin{IEEEeqnarray}{rCl}
% M_{AC}(x) & = & \frac{\sqrt{2}}{3} \nonumber\\
% && \cdot\> \sqrt{(I_{1}(x)-I_{2}(x))^{2}+(I_{2}(x)-I_{3}(x))^{2}+(I_{3}(x)-I_{1}(x))^{2}}
% \label{eq:demod}
% \end{IEEEeqnarray}

% \begin{table}
% \renewcommand{\arraystretch}{1.3}
% \caption{Summary of optical properties of homogeneous tissue phantoms used in this study}
% \label{table:phantoms}
% \centering
% \begin{tabular}{|c|c c|}
% \hline
% Phantom \# & $\mu_{a}$ [mm\textsuperscript{-1}] & $\mu_{s}'$ [mm\textsuperscript{-1}]\\
% \hline
% 1 & 0.0058 & 0.9298\\
% 2 & 0.0062 & 1.1453\\
% 3 & 0.0087 & 0.5801\\
% 4 & 0.0093 & 0.3365\\
% 5 & 0.0119 & 1.0474\\
% 6 & 0.0136 & 1.5534\\
% 7 & 0.0158 & 0.5760\\
% 8 & 0.0190 & 1.2128\\
% 9 & 0.0203 & 1.2387\\
% 10 & 0.0237 & 1.4816\\
% 11 & 0.0291 & 1.8663\\
% 12 & 0.0301 & 0.5304\\
% 13 & 0.0387 & 0.5780\\
% 14 & 0.0416 & 0.9194\\
% 15 & 0.0572 & 0.3176\\
% 16 & 0.1289 & 1.7495\\
% 17 & 0.1565 & 0.9019\\
% 18 & 0.1826 & 0.5755\\
% \hline
% \end{tabular}
% \end{table}

In this study, six tissue-mimicking phantoms were used for training and twelve for testing. We intentionally selected phantoms for training that had optical properties not spanned by the esophagus training samples (highlighted by green ellipses in Fig. \ref{fig:scatter_train}), in order to develop GANPOP networks capable of estimating the widest range of optical properties.

\subsubsection{\textit{In vivo} samples}
To provide the network with \textit{in vivo} samples that were perfused and oxygenated, seven human hands with different levels of pigmentation (Fitzpatrick skin types 1-6) were imaged with SFDI. Two were used for training and five for testing.

\subsubsection{Swine tissue}
Four specimens of upper gastrointestinal tracts that included stomach and esophagus were harvested from four different pigs for \textit{ex vivo} imaging with SFDI. Optical properties of these samples are summarized in Fig. \ref{fig:scatter_test}. Additionally, we imaged a pig colon \textit{in vivo} during a surgery. The live study was performed with approval from Johns Hopkins University Animal Care and Use Committee (ACUC). All swine tissue images were used exclusively for testing optical property prediction.

\subsection{Performance Metric}
Normalized Mean Absolute Error (NMAE) was used to evaluate the performance of different methods, which was calculated using: 

\begin{equation}
NMAE=\frac{\sum_{i=1}^{T}|p_{i}-p_{i,ref}|}{\sum_{i=1}^{T}p_{i,ref}}.
\label{eq:nmae}
\end{equation}

$p_{i}$ and $p_{i,ref}$ are pixel values of predicted and ground-truth data, and $T$ is the total number of pixels. The metric was calculated using SFDI output as ground truth. A smaller NMAE value indicates better performance.

\begin{figure}
\centering
\includegraphics[width=3.5in]{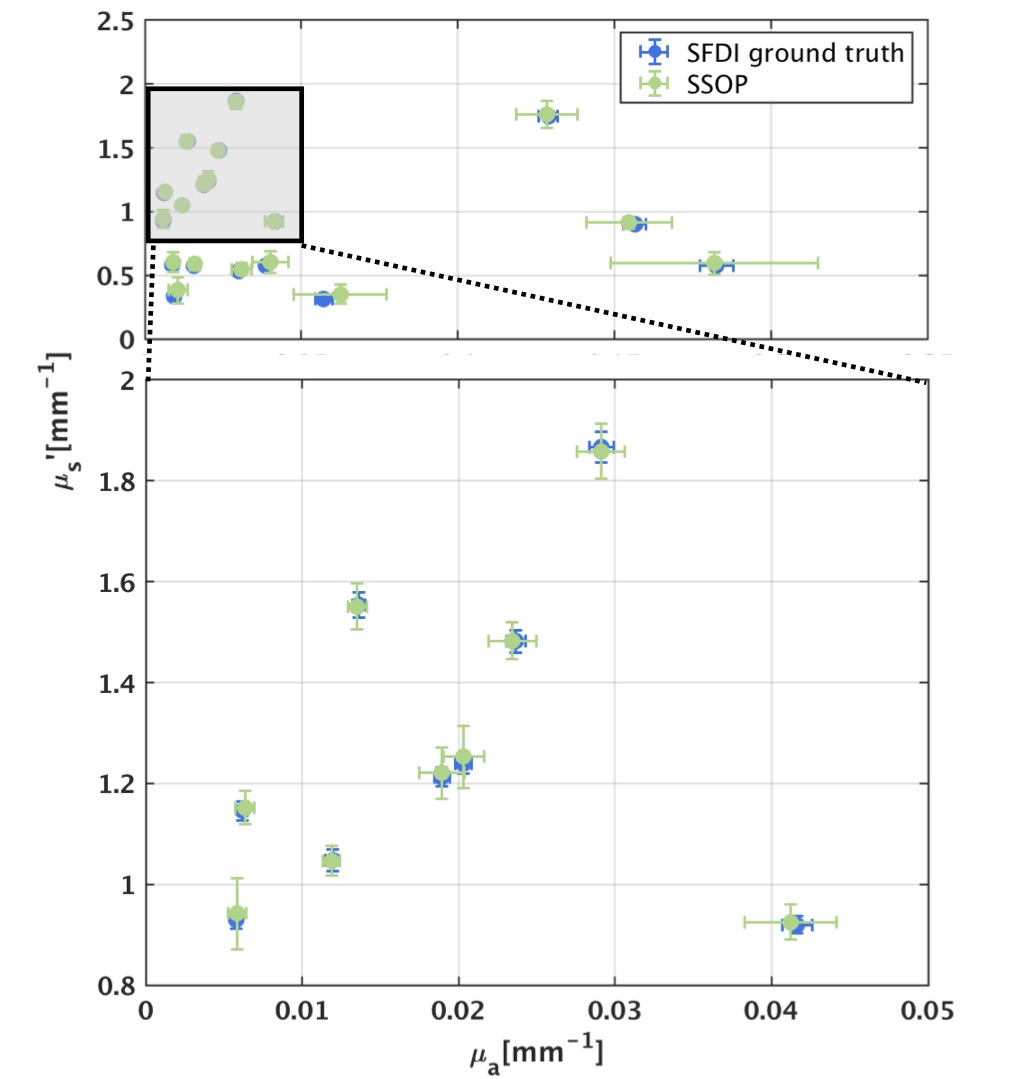}
\caption{Validation of SSOP model-based prediction of optical properties with 18 homogeneous tissue phantoms. SSOP prediction from a single input image demonstrates close agreement with SFDI prediction from six input images.}
\label{fig:ssop_phantoms}
\end{figure}

\section{Results}
\subsection{SSOP validation}
For benchmarking, SSOP was implemented as a model-based counterpart of GANPOP. For independent validation, we applied SSOP to 18 homogeneous tissue phantoms (Fig. \ref{fig:ssop_phantoms}). Each value was calculated as the mean of a 100 $\times$ 100-pixel region of interest (ROI) from the center of the phantom, with error bars showing standard deviations. SSOP demonstrates high accuracy in predicting optical properties of the phantoms, with an average percentage error of 2.35\% for absorption and 2.69\% for reduced scattering.

\subsection{GANPOP test in homogeneous phantoms}
\begin{figure}
\centering
\includegraphics[width=3.5in]{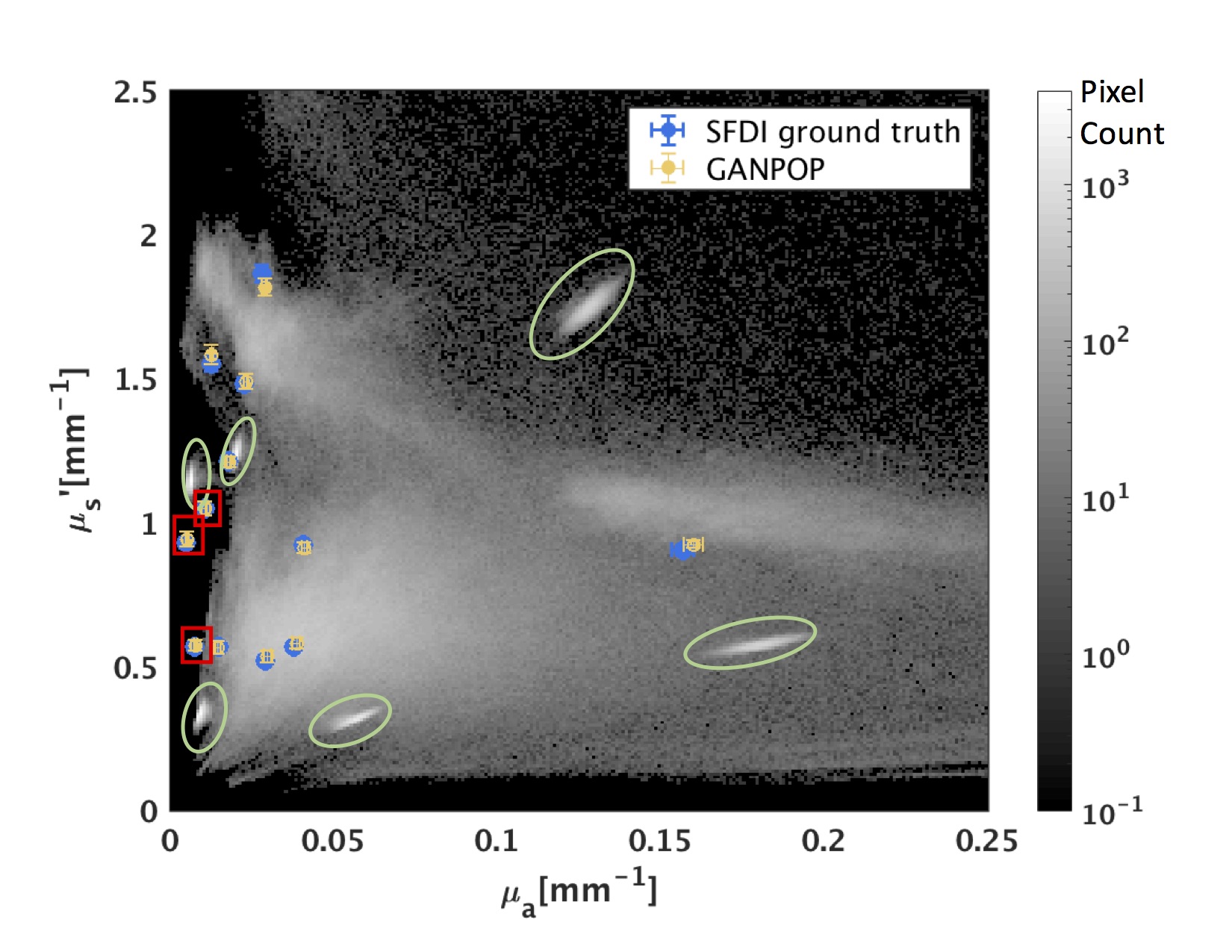}
\caption{Scatter plot showing optical property pairs estimated by GANPOP compared to ground truth from conventional SFDI on 12 tissue phantoms. The 2D histogram in the background illustrates the distribution of training pixels among all optical property pairs, determined by SFDI. Green ellipses indicate dense pixel counts due to homogeneous phantoms used in training. Testing samples in the red box fell outside of the training range but were accurately predicted by GANPOP.}
\label{fig:scatter_train}
\end{figure}

Phantom optical properties predicted by $N_{1}$ are plotted with ground truth in Fig. \ref{fig:scatter_train}. Each optical property reported is the average value of a 100 $\times$ 100 ROI of a homogeneous phantom, with error bars showing standard deviations. On average, GANPOP produced 3.06\% error for absorption and 1.26\% for scattering. The scatter plot in Fig. \ref{fig:scatter_train} is overlaid on a 2D histogram of pixel counts for each ($\mu_{a}$, $\mu_{s}'$) pair used in an example training iteration. Green ellipses indicate training samples from homogeneous phantoms. The three testing results enclosed by red boxes have optical properties outside of the range spanned by the training data but were still reasonably estimated by the GANPOP network. 

\begin{figure}
\centering
\includegraphics[width=3.5in]{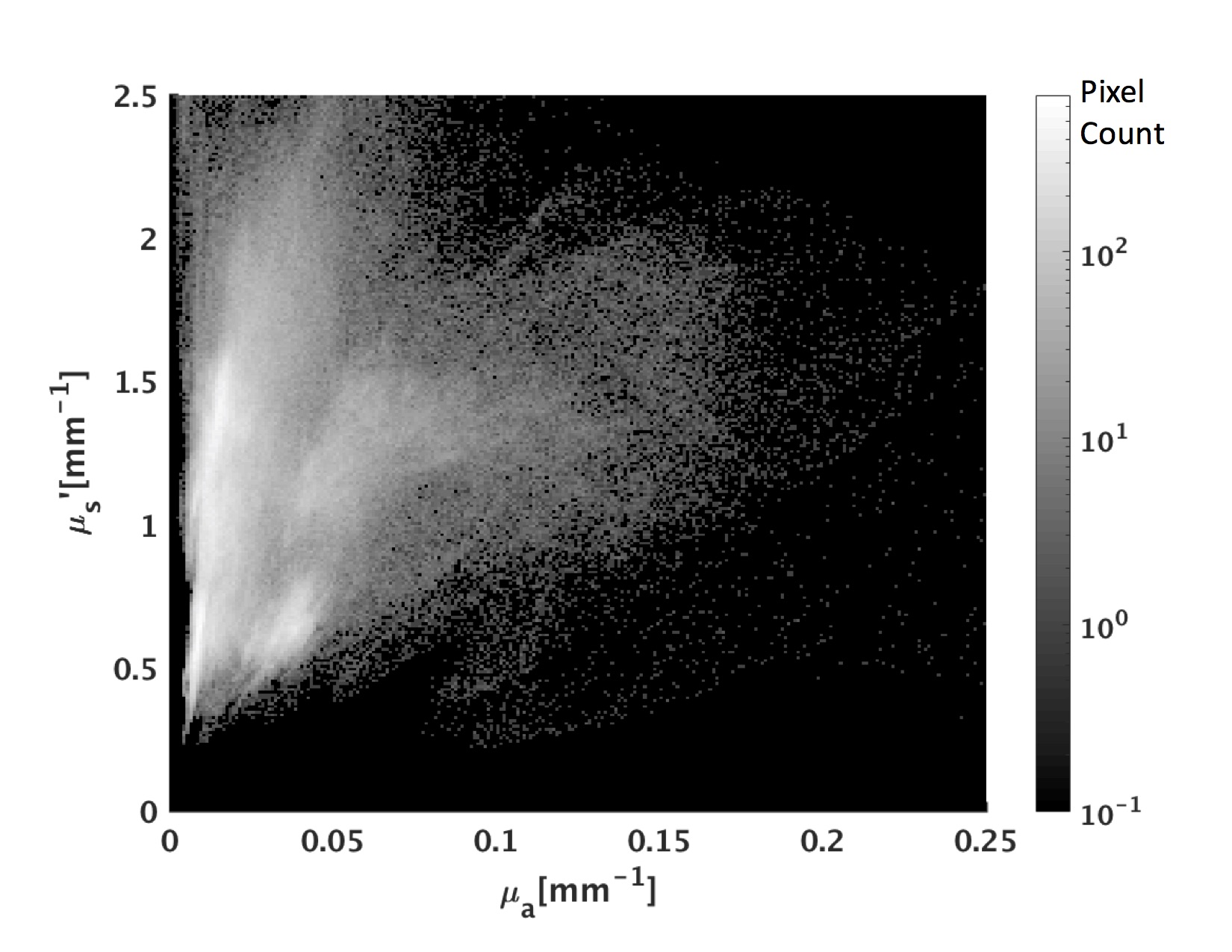}
\caption{Histogram of optical property distribution of testing pixels from pig samples. Compared to training samples, pig tissues tested in this study had, on average, lower absorption coefficients and higher scattering coefficients.}
\label{fig:scatter_test}
\end{figure}

\subsection{GANPOP test on \textit{ex vivo} human esophagus}
\begin{figure}
\centering
\includegraphics[width=3.5in]{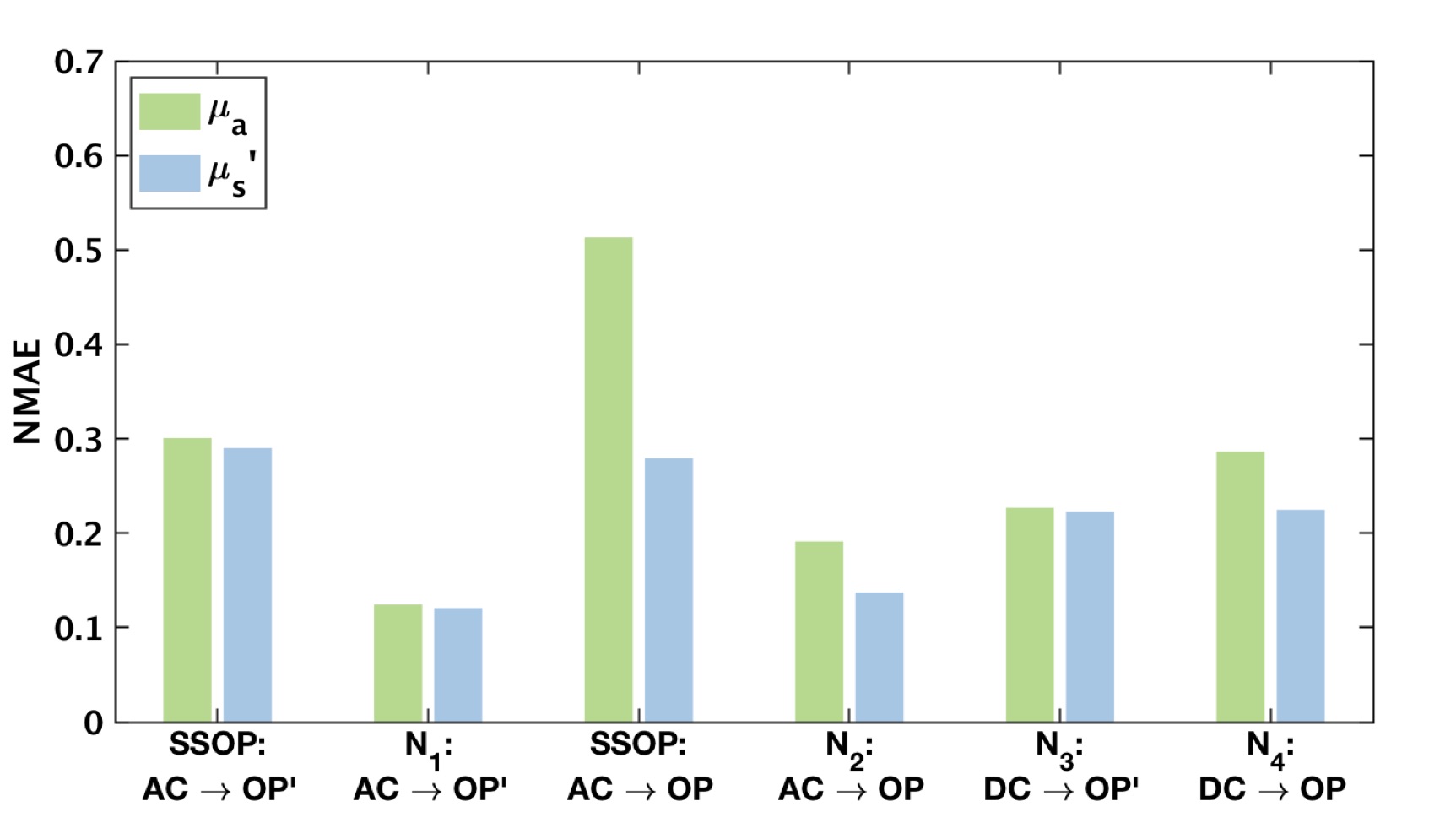}
\caption{Accuracy of GANPOP in predicting optical properties of \textit{ex vivo} human esophagus samples. Average NMAE for absorption (green) and scattering coefficients (blue) are reported for the SSOP technique as well as GANPOP networks. SFDI was used for ground truth. OP and OP' stand for profile-corrected and uncorrected optical properties, respectively.}
\label{fig:nmae_human}
\end{figure}

GANPOP and SSOP were tested on the \textit{ex vivo} human esophagus samples. NMAE scores were calculated for the two testing samples from each of four-fold cross validation iterations, and the average values from the four networks tested on a total of eight samples are reported in Fig. \ref{fig:nmae_human}. Results from $N_{2}$, $N_{4}$, and SSOP are also compared to profilometry-corrected ground truth and shown in the same bar chart. On average, GANPOP produced approximately 58\% higher accuracy with AC input than SSOP. Example optical property maps of a testing sample generated by $N_{1}$ are shown in Fig. \ref{fig:results}(a).

\subsection{GANPOP test on \textit{ex vivo} pig samples}
\begin{figure}
\centering
\includegraphics[width=3.5in]{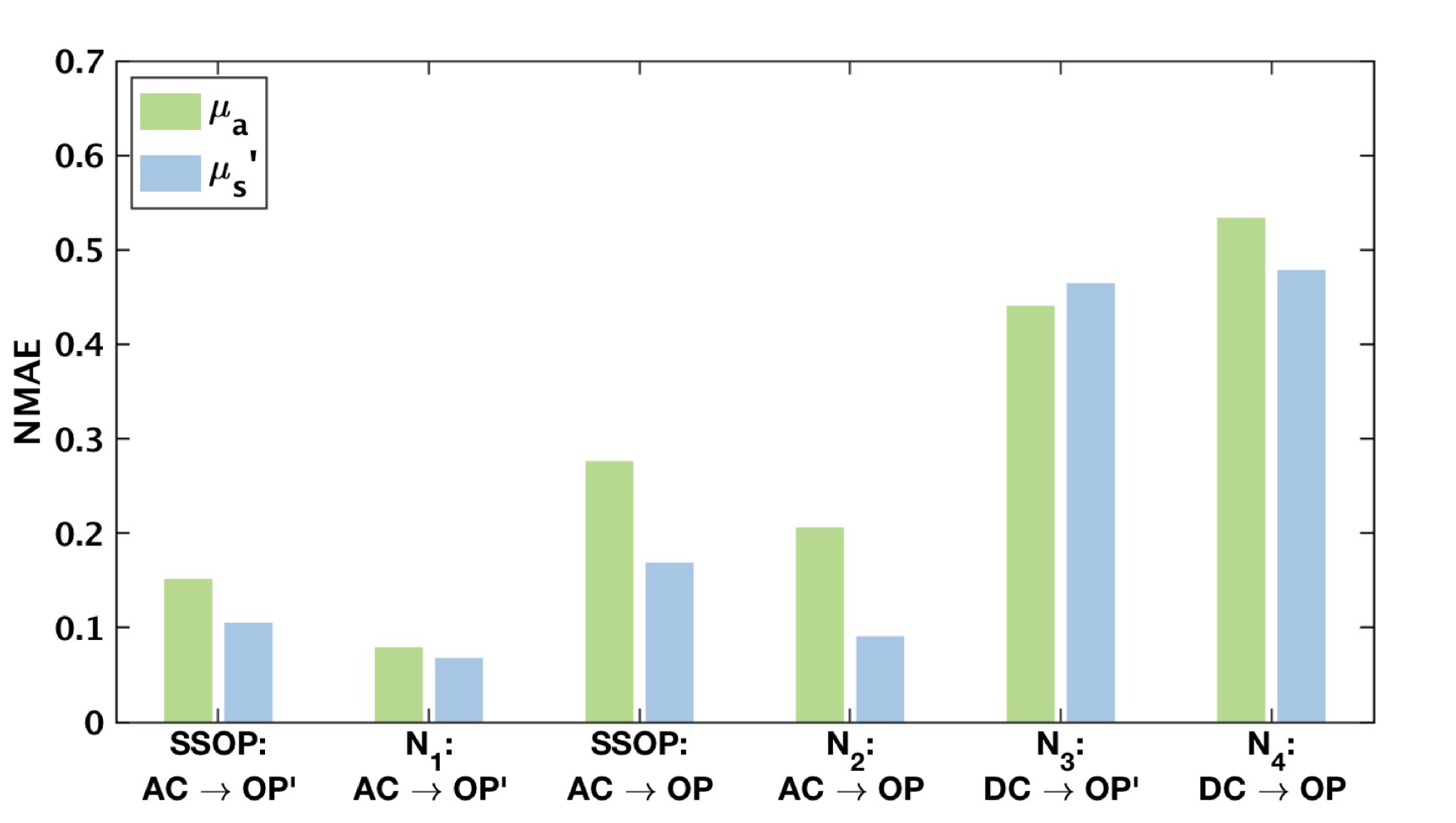}
\caption{Plots of average NMAE for absorption (green) and scattering coefficients (blue) of \textit{ex vivo} pig stomach and esophagus samples. Metrics are calculated against SFDI ground truth. OP and OP' stand for profile-corrected and uncorrected optical properties, respectively.}
\label{fig:nmae_pigs}
\end{figure}

Each of the four GANPOP networks were tested on \textit{ex vivo} esophagus and stomach samples from four pigs. Average NMAE scores for GANPOP and SSOP method were calculated for all eight pig tissue specimens (four esophagi and four stomachs) and are summarized in Fig. \ref{fig:nmae_pigs}. Background regions, which were absorbing paper, were manually masked in the calculation, and the reported scores are the average values of 779,101 tissue pixels. The optical properties of the pig samples are also shown in a 2D histogram in Fig. \ref{fig:scatter_test}. Despite the fact that some testing samples had optical properties not covered by the training set, GANPOP outperforms SSOP in terms of average accuracy and qualitative image quality (Fig. \ref{fig:results}).

\subsection{GANPOP test on \textit{in vivo} pig colon}
\begin{figure}
\centering
\includegraphics[width=3.5in]{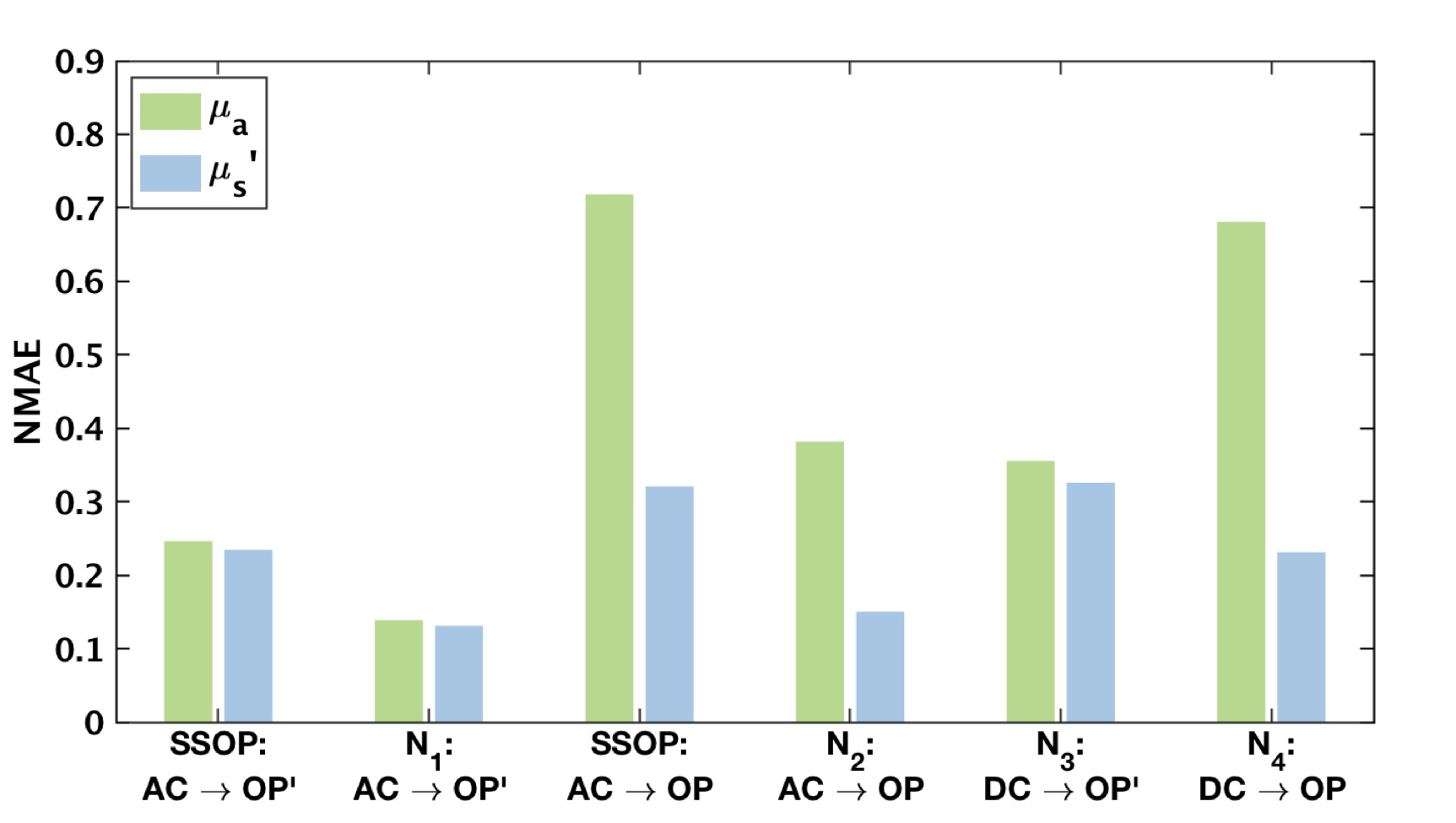}
\caption{Plots of average NMAE for absorption (green) and scattering coefficients (blue) of the \textit{in vivo} pig colon. Metrics are calculated against SFDI ground truth. OP and OP' stand for profile-corrected and uncorrected optical properties, respectively.}
\label{fig:nmae_livepig}
\end{figure}

\begin{figure*}
\centering
\includegraphics[width=7in]{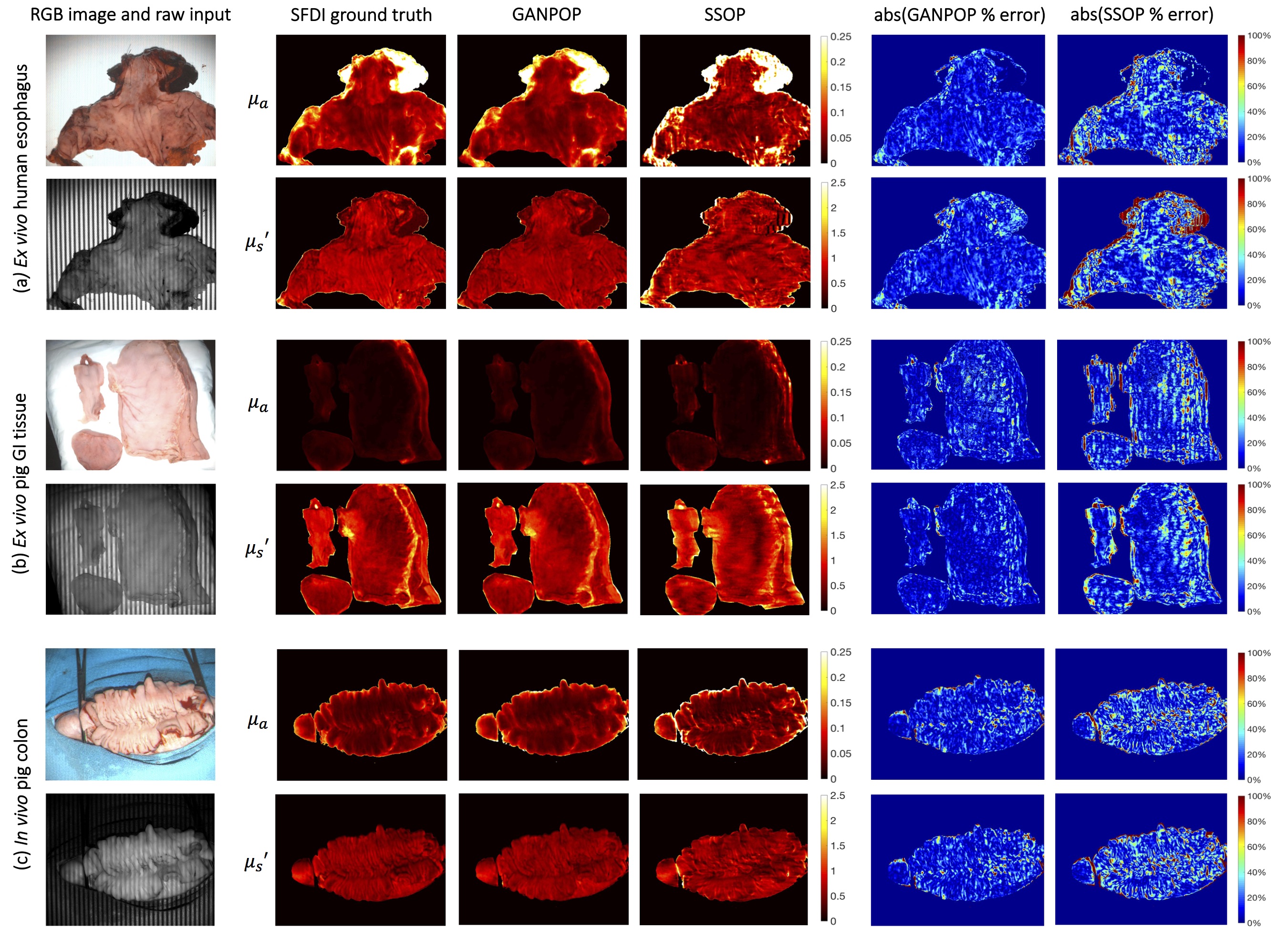}
\caption{Example results for AC input to non-profilometry-corrected optical properties ($N_{1}$). From left to right: RGB image and raw structured illumination image, SFDI ground truth, GANPOP output, SSOP output, percent difference between GANPOP and ground truth, and percent difference between SSOP and ground truth; From top to bottom: (a) \textit{ex vivo} human esophagus, (b) \textit{ex vivo} pig stomach and esophagus, and (c) \textit{in vivo} pig colon.}
\label{fig:results}
\end{figure*}

The networks were additionally tested on an \textit{in vivo} pig colon. Average NMAE scores for GANPOP and SSOP are reported in Fig. \ref{fig:nmae_livepig} as average values of 118,594 pixels. The generated maps are shown in Fig. \ref{fig:results}(c). The proposed technique produces more accurate results than SSOP when compared to both uncorrected and profile-corrected ground truth data.

\begin{table*}
\renewcommand{\arraystretch}{1.2}
 \caption{Performance comparison of the proposed framework against model-based SSOP and other deep learning architectures when tested on profile-uncorrected data ($N_{1}$). Performance is measured in terms of normalized mean absolute error (NMAE).}
 \label{table:comparison}
\begin{center}
\begin{tabular}{|c|c|c|c|c|c|c|c|c|c|c|c|c|c|c|}
\hline
\multirow{2}{*}{Data type} & \multicolumn{2}{c|}{ResNet} & \multicolumn{2}{c|}{UNet} & \multicolumn{2}{c|}{ResNet-UNet} & \multicolumn{2}{c|}{ResNet GAN} & \multicolumn{2}{c|}{UNet GAN} & \multicolumn{2}{c|}{SSOP} &\multicolumn{2}{c|}{\textbf{Proposed}}\\
\cline{2-15}
& $\mu_{a}$ & $\mu_{s}'$ & $\mu_{a}$ & $\mu_{s}'$ & $\mu_{a}$ & $\mu_{s}'$ & $\mu_{a}$ & $\mu_{s}'$ & $\mu_{a}$ & $\mu_{s}'$ & $\mu_{a}$ & $\mu_{s}'$ & $\mu_{a}$ & $\mu_{s}'$\\
\hline
 \textbf{Human esophagus} & 0.227 & 0.143 & 0.161 & 0.153 & 0.203 & 0.140 & 0.232 & 0.156 & 0.176 & 0.165 & 0.301 & 0.290 & \textbf{0.124} & \textbf{0.121}\\
\hline
 \textbf{\textit{In vivo} pig colon} & 0.614 & 0.729 & 0.320 & 0.486 & 0.609 & 0.583 & 0.795 & 0.769 & 0.335 & 0.377 & 0.246 & 0.235 & \textbf{0.139} & \textbf{0.131}\\
\hline
  \textbf{\textit{Ex vivo} pig GI tissue} & 2.954 & 0.175 & 0.344 & 0.378 & 2.842 & 0.177 & 3.138 & 0.175 & 0.574 & 0.410 & 0.152 & 0.106 & \textbf{0.080} & \textbf{0.068}\\
\hline
  \textbf{\textit{In vivo} human hands} & 0.373 & 0.100 & 0.123 & 0.109 & 0.249 & 0.081 & 0.353 & 0.106 & 0.162 & 0.099 & 0.092 & 0.058 & \textbf{0.075} & \textbf{0.055}\\
\hhline{|=|=|=|=|=|=|=|=|=|=|=|=|=|=|=|}
  \textbf{Overall} & 1.042 & 0.287 & 0.237 & 0.281 & 0.976 & 0.245 & 1.129 & 0.301 & 0.312 & 0.263 & 0.198 & 0.172 & \textbf{0.104} & \textbf{0.094}\\
\hline
\end{tabular}
\end{center}
\end{table*}

\subsection{Comparative analysis of existing deep networks}
Several deep learning architectures were explored for the purpose of optical property mapping, including conventional U-Net \cite{Ronneberger_2015} and ResNet \cite{He_2016}, both stand-alone and integrated in a cGAN framework \cite{Isola_2017,Quan2016FusionNetAD}. The NMAE performance of each architecture was compared to GANPOP. All the networks were four-fold cross validated, and the testing dataset included eight \textit{ex vivo} human esophagi, four \textit{ex vivo} pig GI samples, one \textit{in vivo} pig colon, and five \textit{in vivo} hands (Table \ref{table:comparison}).

\section{Discussion}
In this study, we have described a GAN-based technique for end-to-end optical property mapping from single structured and flat-field illumination images. Compared to the original pix2pix paradigm \cite{Isola_2017}, the generator of our model adopted a fusion of U-Net and ResNet architectures for several reasons. First, a fully residual network effectively resolved the issue of vanishing gradients, allowing us to stably train a relatively deep neural network \cite{Quan2016FusionNetAD}. Second, the use of both long and short skip connections enables the network to learn from the structure of the images while preserving both low and high frequency details. The information flow both within and between levels is important for the prediction of optical properties, as demonstrated by the improved performance over a U-Net or ResNet approach. Moreover, as shown in Table \ref{table:comparison}, the inclusion of a discriminator significantly improved the performance of the fusion generator. This was especially apparent in the case for pig data, likely due to this testing tissue differing considerably from the training samples. We hypothesize that the cGAN architecture enforced the similarity between generated images and ground truth while preventing the generator from depending too much on the context of the image. Overall, the GANPOP method outperformed the other deep networks by a significant margin on all data types (Table \ref{table:comparison}). Additionally, we empirically found that a least squares GAN outperformed a conventional GAN when trained for 200 epochs. However, as discussed in \cite{lucic_are_2017}, this improvement could potentially be matched by a conventional GAN with more training. 

Compared to phantom ground truth in Fig. \ref{fig:scatter_train}, GANPOP estimated optical properties with standard deviations on the same order of magnitude as conventional SFDI. Additionally, the GANPOP networks exhibited potential to extrapolate phantom optical properties that were not present in the training samples (highlighted by the red boxes in Fig. \ref{fig:scatter_train}). This provides evidence that these networks have successfully learned the relationship between diffuse reflectance and optical properties, and are able to infer beyond the range of training data.

Fig. \ref{fig:nmae_human}, \ref{fig:nmae_pigs}, and \ref{fig:nmae_livepig} show that GANPOP with AC input consistently outperformed SSOP when tested on these types of data. From Fig. \ref{fig:scatter_test}, it is evident that optical properties of the pig samples differed considerably from those of human esophagi used for training. Nevertheless, GANPOP exhibited more accurate estimation than the model-based SSOP benchmark. Moreover, a single network was trained for estimating both $\mu_{a}$ and $\mu_{s}'$ due to its lower computational cost and potential benefits in learning the relationships between the two parameters in tissues. 

Compared to SSOP, GANPOP optical property maps contain fewer artifacts caused by frequency filtering (Fig. \ref{fig:results}). For both GANPOP and SSOP optical property estimation, a relatively large error is present on the edge of the sample. This is caused by the transition between tissue and the background, which poses problems for SFDI ground truth, and would be less significant for \textit{in vivo} imaging. Artifacts caused by patched input are visible in GANPOP images, which can be reduced by using a larger patch size. However, this was not implemented in our study due to the size and the number of the specimens available for training. In our benchmarking with SSOP, we implemented the first version of the technique, which does not correct for sample height and surface angle variations. Recent developments have enabled these corrections by utilizing a more complex illumination pattern and additional processing steps \cite{van_de_giessen_real-time_2015}. We implemented the original version of SSOP because it allowed comparing identical input images for both SSOP and GANPOP.

In addition to training GANPOP models to estimate optical properties from objects assumed to be flat ($N_{1}$ and $N_{3}$), we also trained networks that directly estimate profilometry-corrected optical properties ($N_{2}$ and $N_{4}$). For the same AC input, these models generated improved results over SSOP when tested on human and pig data. Moreover, when compared against profile-corrected ground truth, they produced 35.7\% less error for $\mu_{a}$ and 44.7\% for $\mu_{s}'$ than did uncorrected GANPOP results from $N_{1}$ and $N_{3}$. This means that GANPOP is capable of inferring surface profile from a single fringe image and adjusting measured diffuse reflectance accordingly. In experiment $N_{3}$ and $N_{4}$, when trained on DC illumination images, the GANPOP model became less accurate. Nevertheless, these networks converged during training, and albeit less accurate, the \textit{ex vivo} human results still produced a lower NMAE than SSOP. Hence, given a sufficiently large training dataset, GANPOP has the potential to enable rapid and accurate wide-field measurements of optical properties from conventional camera systems. This could be useful for applications such as endoscopic imaging of the GI tract, where the range of tissue optical properties is limited and modification of the hardware system is challenging.

In terms of speed, GANPOP requires capturing one sample image instead of six, thus significantly shortening data acquisition time. For optical property extraction, the model developed here without optimization takes approximately 0.04 s to process a 256 $\times$ 256 image on an NVIDIA Tesla P100 GPU. Therefore, this technique has the potential to be applied in real time for fast and accurate optical property mapping. In terms of adaptability, random cropping ensures that our trained models work on any 256 $\times$ 256 patches within the field of view. Additionally, while the models were trained on the same calibration phantom at 660 nm, they could theoretically be applied to other references or wavelengths by scaling the average $M_{DC,ref}$ and $M_{AC,ref}$.

For future work, a more generalizable model could be trained on a wider range of optical properties and imaging geometries, though this would inevitably incur a higher computational cost and necessitate a much larger dataset for training. For example, all input images used here were acquired at an approximately-constant working distance. Incorporating monocular depth estimates into the prediction may enable GANPOP to account for large differences in working distance \cite{mahmood2018unsupervised,Mahmood_2018}. This could be particularly useful for endoscopic screening where constant imaging geometries are difficult to achieve. Having a model trained on images at multiple wavelengths, this technique can be modified to provide critical information in real time, such as tissue oxygenation and metabolism biomarkers. Accuracy in this application may also benefit from training adversarial networks to directly estimate these biomarkers rather than using optical properties as intermediate representations. By similar extension, future research may develop networks to directly estimate disease diagnosis and localization from structured light images. 

\section{Conclusion}
We have proposed a deep learning-based approach to optical property mapping (GANPOP) from single snapshot wide-field images. This model utilizes a conditional Generative Adversarial Network consisting of a generator and a discriminator that are iteratively trained in concert with one another. Using SFDI-determined optical properties as ground truth, GANPOP produces significantly more accurate optical property maps than a model-based SSOP benchmark. Importantly, we have demonstrated that GANPOP can estimate optical properties with conventional flat-field illumination, potentially enabling optical property mapping in endoscopy without modifications for structured illumination. This method lays the foundation for future work in incorporating real-time, high-fidelity optical property mapping and quantitative biomarker imaging into endoscopy and image-guided surgery applications.

% \appendices
% \section{}
% \section{}
% Appendix two

\section*{Acknowledgment}
This work was supported in part with funding from the NIH Trailblazer Award (R21 EB024700).

We would like to thank Dr. Darren Roblyer’s group at Boston University for sharing SFDI software. 

\ifCLASSOPTIONcaptionsoff
  % \newpage
\fi

% \bibliographystyle{IEEEtran}
% \bibliography{ref

\begin{thebibliography}{10}
\providecommand{\url}[1]{#1}
\csname url@samestyle\endcsname
\providecommand{\newblock}{\relax}
\providecommand{\bibinfo}[2]{#2}
\providecommand{\BIBentrySTDinterwordspacing}{\spaceskip=0pt\relax}
\providecommand{\BIBentryALTinterwordstretchfactor}{4}
\providecommand{\BIBentryALTinterwordspacing}{\spaceskip=\fontdimen2\font plus
\BIBentryALTinterwordstretchfactor\fontdimen3\font minus
  \fontdimen4\font\relax}
\providecommand{\BIBforeignlanguage}[2]{{%
\expandafter\ifx\csname l@#1\endcsname\relax
\typeout{** WARNING: IEEEtran.bst: No hyphenation pattern has been}%
\typeout{** loaded for the language `#1'. Using the pattern for}%
\typeout{** the default language instead.}%
\else
\language=\csname l@#1\endcsname
\fi
#2}}
\providecommand{\BIBdecl}{\relax}
\BIBdecl

\bibitem{doi:10.1146/annurev.physchem.47.1.555}
R.~Richards-Kortum and E.~Sevick-Muraca, ``Quantitative optical spectroscopy
  for tissue diagnosis,'' \emph{Annual Review of Physical Chemistry}, vol.~47,
  no.~1, pp. 555--606, 1996.

\bibitem{drezek2003light}
R.~A. Drezek, M.~Guillaud, T.~G. Collier, I.~Boiko, A.~Malpica, C.~E. MacAulay,
  M.~Follen, and R.~R. Richards-Kortum, ``Light scattering from cervical cells
  throughout neoplastic progression: influence of nuclear morphology, dna
  content, and chromatin texture,'' \emph{Journal of biomedical optics},
  vol.~8, no.~1, pp. 7--17, 2003.

\bibitem{maloney2018review}
B.~W. Maloney, D.~M. McClatchy, B.~W. Pogue, K.~D. Paulsen, W.~A. Wells, and
  R.~J. Barth, ``Review of methods for intraoperative margin detection for
  breast conserving surgery,'' \emph{Journal of biomedical optics}, vol.~23,
  no.~10, p. 100901, 2018.

\bibitem{mourant2000light}
J.~R. Mourant, M.~Canpolat, C.~Brocker, O.~Esponda-Ramos, T.~M. Johnson,
  A.~Matanock, K.~Stetter, and J.~P. Freyer, ``Light scattering from cells: the
  contribution of the nucleus and the effects of proliferative status,''
  \emph{Journal of biomedical optics}, vol.~5, no.~2, pp. 131--138, 2000.

\bibitem{Steelman:19}
Z.~A. Steelman, D.~S. Ho, K.~K. Chu, and A.~Wax, ``Light-scattering methods for
  tissue diagnosis,'' \emph{Optica}, vol.~6, no.~4, pp. 479--489, Apr 2019.

\bibitem{Lin2011}
A.~J. Lin, M.~A. Koike, K.~N. Green, J.~G. Kim, A.~Mazhar, T.~B. Rice, F.~M.
  LaFerla, and B.~J. Tromberg, ``Spatial frequency domain imaging of intrinsic
  optical property contrast in a mouse model of alzheimer's disease,''
  \emph{Annals of Biomedical Engineering}, vol.~39, no.~4, pp. 1349--1357, Apr
  2011.

\bibitem{Shah4420}
N.~Shah, A.~Cerussi, C.~Eker, J.~Espinoza, J.~Butler, J.~Fishkin, R.~Hornung,
  and B.~Tromberg, ``Noninvasive functional optical spectroscopy of human
  breast tissue,'' \emph{Proceedings of the National Academy of Sciences},
  vol.~98, no.~8, pp. 4420--4425, 2001.

\bibitem{dognitz1998determination}
N.~D{\"o}gnitz and G.~Wagni{\`e}res, ``Determination of tissue optical
  properties by steady-state spatial frequency-domain reflectometry,''
  \emph{Lasers in medical science}, vol.~13, no.~1, pp. 55--65, 1998.

\bibitem{cuccia_quantitation_2009}
D.~J. Cuccia, F.~Bevilacqua, A.~J. Durkin, F.~R. Ayers, and B.~J. Tromberg,
  ``\BIBforeignlanguage{eng}{Quantitation and mapping of tissue optical
  properties using modulated imaging},'' \emph{\BIBforeignlanguage{eng}{Journal
  of Biomedical Optics}}, vol.~14, no.~2, p. 024012, Apr. 2009.

\bibitem{pharaon_early_2010}
M.~R. Pharaon, T.~Scholz, S.~Bogdanoff, D.~Cuccia, A.~J. Durkin, D.~B. Hoyt,
  and G.~R.~D. Evans, ``\BIBforeignlanguage{eng}{Early detection of complete
  vascular occlusion in a pedicle flap model using quantitative [corrected]
  spectral imaging},'' \emph{\BIBforeignlanguage{eng}{Plastic and
  Reconstructive Surgery}}, vol. 126, no.~6, pp. 1924--1935, Dec. 2010.

\bibitem{gioux_first--human_2011}
S.~Gioux, A.~Mazhar, B.~T. Lee, S.~J. Lin, A.~M. Tobias, D.~J. Cuccia,
  A.~Stockdale, R.~Oketokoun, Y.~Ashitate, E.~Kelly, M.~Weinmann, N.~J. Durr,
  L.~A. Moffitt, A.~J. Durkin, B.~J. Tromberg, and J.~V. Frangioni,
  ``First-in-human pilot study of a spatial frequency domain oxygenation
  imaging system,'' \emph{Journal of Biomedical Optics}, vol.~16, no.~8, 2011.

\bibitem{kaiser_noninvasive_2011}
M.~Kaiser, A.~Yafi, M.~Cinat, B.~Choi, and A.~J. Durkin,
  ``\BIBforeignlanguage{eng}{Noninvasive assessment of burn wound severity
  using optical technology: a review of current and future modalities},''
  \emph{\BIBforeignlanguage{eng}{Burns: Journal of the International Society
  for Burn Injuries}}, vol.~37, no.~3, pp. 377--386, May 2011.

\bibitem{WEINKAUF2019555}
C.~Weinkauf, A.~Mazhar, K.~Vaishnav, A.~A. Hamadani, D.~J. Cuccia, and D.~G.
  Armstrong, ``Near-instant noninvasive optical imaging of tissue perfusion for
  vascular assessment,'' \emph{Journal of Vascular Surgery}, vol.~69, no.~2,
  pp. 555 -- 562, 2019.

\bibitem{doi:10.1002/lsm.22692}
A.~Yafi, F.~K. Muakkassa, T.~Pasupneti, J.~Fulton, D.~J. Cuccia, A.~Mazhar,
  K.~N. Blasiole, and E.~N. Mostow, ``Quantitative skin assessment using
  spatial frequency domain imaging (sfdi) in patients with or at high risk for
  pressure ulcers,'' \emph{Lasers in Surgery and Medicine}, vol.~49, no.~9, pp.
  827--834, 2017.

\bibitem{angelo_real-time_2017}
J.~P. Angelo, M.~van~de Giessen, and S.~Gioux, ``Real-time endoscopic optical
  properties imaging,'' \emph{Biomedical Optics Express}, vol.~8, no.~11, pp.
  5113--5126, Oct. 2017.

\bibitem{nandy_label-free_2018}
S.~Nandy, W.~Chapman, R.~Rais, I.~González, D.~Chatterjee, M.~Mutch, and
  Q.~Zhu, ``\BIBforeignlanguage{en}{Label-free quantitative optical assessment
  of human colon tissue using spatial frequency domain imaging},''
  \emph{\BIBforeignlanguage{en}{Techniques in Coloproctology}}, vol.~22, no.~8,
  pp. 617--621, Aug. 2018.

\bibitem{vervandier_single_2013}
J.~Vervandier and S.~Gioux, ``Single snapshot imaging of optical properties,''
  \emph{Biomedical Optics Express}, vol.~4, no.~12, pp. 2938--2944, Nov. 2013.

\bibitem{suzuki_overview_2017}
K.~Suzuki, ``\BIBforeignlanguage{eng}{Overview of deep learning in medical
  imaging},'' \emph{\BIBforeignlanguage{eng}{Radiological Physics and
  Technology}}, vol.~10, no.~3, pp. 257--273, Sep. 2017.

\bibitem{7404017}
H.~{Shin}, H.~R. {Roth}, M.~{Gao}, L.~{Lu}, Z.~{Xu}, I.~{Nogues}, J.~{Yao},
  D.~{Mollura}, and R.~M. {Summers}, ``Deep convolutional neural networks for
  computer-aided detection: Cnn architectures, dataset characteristics and
  transfer learning,'' \emph{IEEE Transactions on Medical Imaging}, vol.~35,
  no.~5, pp. 1285--1298, May 2016.

\bibitem{7426826}
N.~{Tajbakhsh}, J.~Y. {Shin}, S.~R. {Gurudu}, R.~T. {Hurst}, C.~B. {Kendall},
  M.~B. {Gotway}, and J.~{Liang}, ``Convolutional neural networks for medical
  image analysis: Full training or fine tuning?'' \emph{IEEE Transactions on
  Medical Imaging}, vol.~35, no.~5, pp. 1299--1312, May 2016.

\bibitem{goodfellow_generative_2014}
I.~Goodfellow, J.~Pouget-Abadie, M.~Mirza, B.~Xu, D.~Warde-Farley, S.~Ozair,
  A.~Courville, and Y.~Bengio, ``Generative {Adversarial} {Nets},'' in
  \emph{Advances in {Neural} {Information} {Processing} {Systems} 27},
  Z.~Ghahramani, M.~Welling, C.~Cortes, N.~D. Lawrence, and K.~Q. Weinberger,
  Eds.\hskip 1em plus 0.5em minus 0.4em\relax Curran Associates, Inc., 2014,
  pp. 2672--2680.

\bibitem{mirza2014conditional}
M.~Mirza and S.~Osindero, ``Conditional generative adversarial nets,'' 2014.

\bibitem{Isola_2017}
P.~Isola, J.-Y. Zhu, T.~Zhou, and A.~A. Efros, ``Image-to-image translation
  with conditional adversarial networks,'' \emph{2017 IEEE Conference on
  Computer Vision and Pattern Recognition (CVPR)}, Jul 2017.

\bibitem{8662628}
A.~{Diaz-Pinto}, A.~{Colomer}, V.~{Naranjo}, S.~{Morales}, Y.~{Xu}, and A.~F.
  {Frangi}, ``Retinal image synthesis and semi-supervised learning for glaucoma
  assessment,'' \emph{IEEE Transactions on Medical Imaging}, pp. 1--1, 2019.

\bibitem{8340157}
Q.~{Yang}, P.~{Yan}, Y.~{Zhang}, H.~{Yu}, Y.~{Shi}, X.~{Mou}, M.~K. {Kalra},
  Y.~{Zhang}, L.~{Sun}, and G.~{Wang}, ``Low-dose ct image denoising using a
  generative adversarial network with wasserstein distance and perceptual
  loss,'' \emph{IEEE Transactions on Medical Imaging}, vol.~37, no.~6, pp.
  1348--1357, June 2018.

\bibitem{8233175}
G.~{Yang}, S.~{Yu}, H.~{Dong}, G.~{Slabaugh}, P.~L. {Dragotti}, X.~{Ye},
  F.~{Liu}, S.~{Arridge}, J.~{Keegan}, Y.~{Guo}, and D.~{Firmin}, ``Dagan: Deep
  de-aliasing generative adversarial networks for fast compressed sensing mri
  reconstruction,'' \emph{IEEE Transactions on Medical Imaging}, vol.~37,
  no.~6, pp. 1310--1321, June 2018.

\bibitem{Swartling:03}
J.~Swartling, J.~S. Dam, and S.~Andersson-Engels, ``Comparison of spatially and
  temporally resolved diffuse-reflectance measurement systems for determination
  of biomedical optical properties,'' \emph{Appl. Opt.}, vol.~42, no.~22, pp.
  4612--4620, Aug 2003.

\bibitem{doi:10.1063/1.3116135}
J.~R. Weber, D.~J. Cuccia, A.~J. Durkin, and B.~J. Tromberg, ``Noncontact
  imaging of absorption and scattering in layered tissue using spatially
  modulated structured light,'' \emph{Journal of Applied Physics}, vol. 105,
  no.~10, p. 102028, 2009.

\bibitem{palmer_quantitative_2009}
G.~M. Palmer, R.~J. Viola, T.~Schroeder, P.~S. Yarmolenko, M.~W. Dewhirst, and
  N.~Ramanujam, ``Quantitative {Diffuse} {Reflectance} and {Fluorescence}
  {Spectroscopy}: {A} {Tool} to {Monitor} {Tumor} {Physiology} {In} {Vivo},''
  \emph{Journal of biomedical optics}, vol.~14, no.~2, p. 024010, 2009.

\bibitem{7859372}
G.~{Jones}, N.~T. {Clancy}, Y.~{Helo}, S.~{Arridge}, D.~S. {Elson}, and
  D.~{Stoyanov}, ``Bayesian estimation of intrinsic tissue oxygenation and
  perfusion from rgb images,'' \emph{IEEE Transactions on Medical Imaging},
  vol.~36, no.~7, pp. 1491--1501, July 2017.

\bibitem{van_de_giessen_real-time_2015}
M.~van~de Giessen, J.~P. Angelo, and S.~Gioux, ``Real-time, profile-corrected
  single snapshot imaging of optical properties,'' \emph{Biomedical Optics
  Express}, vol.~6, no.~10, pp. 4051--4062, Sep. 2015.

\bibitem{ml_op}
S.~G. Swapnesh~Panigrahi, ``Machine learning approach for rapid and accurate
  estimation of optical properties using spatial frequency domain imaging,''
  \emph{Journal of Biomedical Optics}, vol.~24, no.~7, pp. 1 -- 6 -- 6, 2018.

\bibitem{zhao2018deep}
Y.~Zhao, Y.~Deng, F.~Bao, H.~Peterson, R.~Istfan, and D.~Roblyer, ``Deep
  learning model for ultrafast multifrequency optical property extractions for
  spatial frequency domain imaging,'' \emph{Optics letters}, vol.~43, no.~22,
  pp. 5669--5672, 2018.

\bibitem{gioux_three-dimensional_2009}
S.~Gioux, A.~Mazhar, D.~J. Cuccia, A.~J. Durkin, B.~J. Tromberg, and J.~V.
  Frangioni, ``Three-{Dimensional} {Surface} {Profile} {Intensity} {Correction}
  for {Spatially}-{Modulated} {Imaging},'' \emph{Journal of biomedical optics},
  vol.~14, no.~3, p. 034045, 2009.

\bibitem{martinelli_analysis_2011}
M.~Martinelli, A.~Gardner, D.~Cuccia, C.~Hayakawa, J.~Spanier, and
  V.~Venugopalan, ``\BIBforeignlanguage{EN}{Analysis of single {Monte} {Carlo}
  methods for prediction of reflectance from turbid media},''
  \emph{\BIBforeignlanguage{EN}{Optics Express}}, vol.~19, no.~20, pp.
  19\,627--19\,642, Sep. 2011.

\bibitem{chen2018rethinking}
R.~Chen, F.~Mahmood, A.~Yuille, and N.~J. Durr, ``Rethinking monocular depth
  estimation with adversarial training,'' 2018.

\bibitem{Ronneberger_2015}
O.~Ronneberger, P.~Fischer, and T.~Brox, ``U-net: Convolutional networks for
  biomedical image segmentation,'' \emph{Medical Image Computing and
  Computer-Assisted Intervention – MICCAI 2015}, p. 234–241, 2015.

\bibitem{He_2016}
K.~He, X.~Zhang, S.~Ren, and J.~Sun, ``Deep residual learning for image
  recognition,'' \emph{2016 IEEE Conference on Computer Vision and Pattern
  Recognition (CVPR)}, Jun 2016.

\bibitem{Quan2016FusionNetAD}
T.~M. Quan, D.~G.~C. Hildebrand, and W.-K. Jeong, ``Fusionnet: A deep fully
  residual convolutional neural network for image segmentation in
  connectomics,'' \emph{CoRR}, vol. abs/1612.05360, 2016.

\bibitem{miyato2018cgans}
T.~Miyato and M.~Koyama, ``cgans with projection discriminator,'' 2018.

\bibitem{Mao_2017}
X.~Mao, Q.~Li, H.~Xie, R.~Y. Lau, Z.~Wang, and S.~P. Smolley, ``Least squares
  generative adversarial networks,'' \emph{2017 IEEE International Conference
  on Computer Vision (ICCV)}, Oct 2017.

\bibitem{kingma2014adam}
D.~P. Kingma and J.~Ba, ``Adam: A method for stochastic optimization,'' 2014.

\bibitem{vanhoucke_improving_2011}
V.~Vanhoucke, A.~Senior, and M.~Z. Mao, ``Improving the speed of neural
  networks on {CPUs},'' in \emph{Deep {Learning} and {Unsupervised} {Feature}
  {Learning} {Workshop}, {NIPS} 2011}, 2011.

\bibitem{sweer2019wide}
J.~A. Sweer, K.~Salimian, T.~Chen, R.~J. Battafarano, and N.~J. Durr,
  ``Wide-field optical property mapping and structured light imaging of the
  esophagus with spatial frequency domain imaging,'' \emph{Journal of
  biophotonics}, p. e201900005, 2019.

\bibitem{mazhar_wavelength_2010}
A.~Mazhar, S.~Dell, D.~J. Cuccia, S.~Gioux, A.~J. Durkin, J.~V. Frangioni, and
  B.~J. Tromberg, ``\BIBforeignlanguage{eng}{Wavelength optimization for rapid
  chromophore mapping using spatial frequency domain imaging},''
  \emph{\BIBforeignlanguage{eng}{Journal of Biomedical Optics}}, vol.~15,
  no.~6, p. 061716, Dec. 2010.

\bibitem{ayers_fabrication_2008}
F.~Ayers, A.~Grant, D.~Kuo, D.~J. Cuccia, and A.~J. Durkin, ``Fabrication and
  characterization of silicone-based tissue phantoms with tunable optical
  properties in the visible and near infrared domain,'' in \emph{Design and
  {Performance} {Validation} of {Phantoms} {Used} in {Conjunction} with
  {Optical} {Measurements} of {Tissue}}, vol. 6870.\hskip 1em plus 0.5em minus
  0.4em\relax International Society for Optics and Photonics, Feb. 2008, p.
  687007.

\bibitem{lucic_are_2017}
M.~Lucic, K.~Kurach, M.~Michalski, S.~Gelly, and O.~Bousquet, ``Are {GANs}
  {Created} {Equal}? {A} {Large}-{Scale} {Study},'' \emph{arXiv:1711.10337 [cs,
  stat]}, Nov. 2017, arXiv: 1711.10337.

\bibitem{mahmood2018unsupervised}
F.~Mahmood, R.~Chen, and N.~J. Durr, ``Unsupervised reverse domain adaptation
  for synthetic medical images via adversarial training,'' \emph{IEEE
  transactions on medical imaging}, vol.~37, no.~12, pp. 2572--2581, 2018.

\bibitem{Mahmood_2018}
F.~Mahmood, R.~Chen, S.~Sudarsky, D.~Yu, and N.~J. Durr, ``Deep learning with
  cinematic rendering: fine-tuning deep neural networks using photorealistic
  medical images,'' \emph{Physics in Medicine {\&} Biology}, vol.~63, no.~18,
  p. 185012, sep 2018.

\end{thebibliography}


\end{document}